\documentclass{birkjour}
 \theoremstyle{definition}
 
 \theoremstyle{remark}

 \numberwithin{equation}{section}

\begin{document}

%
%
%
%
%
%
%
%
%

\title{Stability of a class of entropies based on fractional calculus}

\author[Rui A. C. Ferreira]{Rui A. C. Ferreira}

\address{%
Grupo F\'isica-Matem\'atica, Faculdade de Ci\^encias, Universidade de Lisboa,\\
Av. Prof. Gama Pinto, 2, 1649-003, Lisboa, Portugal.} 

\email{raferreira@fc.ul.pt}

\thanks{The author was supported by the ``Funda\c{c}\~{a}o para a Ci\^encia e a Tecnologia (FCT)" through the program ``Stimulus of Scientific Employment, Individual Support-2017 Call" with reference CEECIND/00640/2017.}
\subjclass{82A05, 94A17.}

\keywords{Fractional derivatives, stability, entropy.}


\begin{abstract}
In this work we argue about the Lesche stability of some systems, that are motivated by the use of fractional derivatives.
\end{abstract}

\maketitle
\section{Preamble}

This text is primarily motivated by the work of Ubriaco \cite{MR2542685}, namely, with the proof of the Lesche stability of a certain entropic functional presented therein. Specifically, based on the concept of the Liouville--fractional derivative, the author defined the following entropy (see \cite[Section 3]{MR2542685})
\begin{equation}\label{UE}
    S(p)=\sum_{i=1}^np_i(-\ln(p_i))^\alpha,\quad 0<\alpha\leq 1.
\end{equation}

In 1982, B. Lesche \cite{MR657521} introduced a way of measuring the stability of a system (see also \cite{MR1935190}). Concretely, a system $S$ over a probability distribution $P=\{p_1,\ldots,p_n\}$ with $n\geq 2$ is said to be \emph{Lesche stable} iff
$$\forall\varepsilon>0\ \exists \delta>0\ s.t.\ \forall n,\ \forall p,p'\in P:\|p-p'\|_1\leq \delta\Rightarrow\frac{|S(p)-S(p')|}{S_{\max}}<\varepsilon.$$

In \cite[Section 4.1]{MR2542685} the author \emph{showed} that \eqref{UE} is Lesche stable, using the main result of \cite{MR2098048}. However his proof has an inconsistency, as we shall now explain: having in mind the notation of \cite{MR2098048}, Ubriaco defined in \cite[formula (30)]{MR2542685} the following function,
$$f^{-1}(t)=(-\ln{(t)})^\alpha-\alpha(-\ln(t))^{\alpha-1},\quad \alpha\in (0,1],\ t\in(0,1),$$
which is nothing else but the derivative of $t(-\ln{(t)})^\alpha$. We see that $f^{-1}(1)=\lim_{t\rightarrow 1^-}f^{-1}(t)=-\infty$. Now, the departure formula for the general entropic expression in \cite{MR2098048} is
$$S(p)=\sum_{i=1}^n\left[p_if^{-1}(p_i)-\int_{f^{-1}(0)}^{f^{-1}(p_i)}f(t)dt\right]+\int_{f^{-1}(0)}^{f^{-1}(1)}f(t)dt-f^{-1}(1),$$
where $f$ is a monotonically decreasing function with range $(0,1)$. Therefore, for $p_i<1$, we see that $S(p)\geq +\infty$ which is just nonsense. We also note that the authors assumed differentiability of $f^{-1}$ in order to obtain \cite[formula (6)]{MR2098048}, which will be used somewhere after in their analysis. Summing up, we cannot conclude from the reasoning in \cite[Section 4.1]{MR2542685} that \eqref{UE} is Lesche stable. However, we can conclude it from the results in \cite{MR2525462}. Indeed, from \cite[Example 3]{MR2525462} we know that, defining $g(x)=x^\alpha$ for $x\in[0,\infty)$, then $g(0)=0$ and the function $xg(-\ln{(x)})$ is concave. Therefore, the entropy \eqref{UE} is Lesche stable.

In the next section we will introduce a function that generalizes the entropy given by \eqref{UE}. Some properties of it are presented as well. 

\section{A novel entropic function}

We use the same motivation as in \cite{MR2542685} to define a new entropy, but now by considering the \emph{tempered fractional derivative}.

Let us start by introducing the concept (see e.g. \cite{MR3342453}). For $0<\sigma< 1$ and $\lambda\geq 0$, the Liouville tempered fractional derivative of order $\sigma$ is defined by
\begin{equation}\label{trld}
  D^{\sigma,\lambda}u(t)=\frac{e^{-\lambda t}}{\Gamma(1-\sigma)}\frac{d}{dt}
  \int_{-\infty}^t(t-s)^{-\sigma}e^{\lambda s}u(s)ds.
\end{equation}
For $\lambda=0$ the operator \eqref{trld} is just the Liouville fractional derivative used in \cite[formula (16)]{MR2542685}. Now we wish to calculate \eqref{trld} at $t=-1$ with $u(s)=e^{-s\ln(p_i)},\ p_i<1$ (see \cite[Section 3]{MR2542685}). We have,
\begin{align*}
    \frac{d}{dt}
  \int_{-\infty}^t (t-s)^{-\sigma}e^{\lambda s}e^{-s\ln(p_i)}ds&=\frac{d}{dt}e^{t(\lambda-\ln(p_i))}
  \int_{0}^\infty u^{-\sigma}e^{-u(\lambda-\ln(p_i))}du\\
  &=\frac{d}{dt}e^{t(\lambda-\ln(p_i))}(\lambda-\ln(p_i))^{\sigma-1}\Gamma(1-\sigma)\\
  &=e^{t(\lambda-\ln(p_i))}(\lambda-\ln(p_i))^{\sigma}\Gamma(1-\sigma),
\end{align*}
where we used the definition of the gamma function $\Gamma(t)=\int_0^\infty x^{t-1}e^{-x}dx$. Therefore,
\begin{align*}
D^{\sigma,\lambda}(s\to e^{-s\ln(p_i)})(-1)&=\frac{e^{-\lambda (-1)}}{\Gamma(1-\sigma)}e^{-1(\lambda-\ln(p_i))}(\lambda-\ln(p_i))^{\sigma}\Gamma(1-\sigma)\\
&=p_i(\lambda-\ln(p_i))^\sigma,\quad p_i<1.
\end{align*}

For a probability distribution $P=\{p_1,\ldots,p_n\}$ with $n\geq 2$, we postulate the following two parameter entropy:

\begin{equation}\label{novaEnt}
    S_{\sigma,\lambda}(p)=\sum_{i=1}^n p_i\left[(\lambda-\ln(p_i))^\sigma-\lambda^\sigma\right],\quad\sigma\in(0,1],\ \lambda\geq 0.
\end{equation}
Note that, for $\sigma=1$, we are adopting the usual convention that $x\ln(x)|_{x=0}$.
It is clear that $S_{\sigma,0}$ reduces to the entropy defined by Ubriaco in \eqref{UE}. The entropy defined above satisfies the following three Shannon--Khinchin axioms: let $\Delta_n=\{(p_1,...,p_n)\in \mathbb{R}^n \, | \, p_i\geq 0, \, p_1+...+p_n=1\}$. Then,

\begin{enumerate}
    \item $S_n$   is nonnegative and  continuous  on $\Delta_n$. 
\item For all $(p_1,...,p_n)\in \Delta_n$: \,  $S_n(p_1,...,p_n) \leq S_n(1/n,...,1/n)$. 
\item  For all $(p_1,...,p_n)\in \Delta_n$:  \, $S_{n+1}(p_1,...,p_n,0)=S_n(p_1,...,p_n)$.
\end{enumerate}
Essentially that is due to the fact that the function defined by $$f_{\sigma,\lambda}(x)=x\left[(\lambda-\ln(x))^\sigma-\lambda^\sigma\right],\quad x\in[0,1],$$ 
where we understand for the value $f(0)=\lim_{x\to 0^+}x\left[(\lambda-\ln(x))^\sigma-\lambda^\sigma\right]=0$, is concave. Moreover, defining $g(x)=(\lambda+x)^\sigma-\lambda^\sigma,\ x\geq 0$, we get that $g(0)=0$ and $g$ is concave. Therefore, using again \cite[Example 3]{MR2525462}, we conclude that $S_{\sigma,\lambda}$ given by \eqref{novaEnt} is Lesche stable\footnote{We note that, for $\lambda>0$, the function $f_{\sigma,\lambda}(x)$ is differentiable at $x=1$. Therefore, if we define by $f^{-1}(x)=\frac{d}{dx}x\left[(\lambda-\ln(x))^\sigma-\lambda^\sigma\right]$, then we could actually use the results of \cite{MR2098048} to conclude that \eqref{novaEnt} is Lesche stable.}.

We would like to end this work remarking that $S_{\sigma,\lambda}(p)\leq S_{\sigma,0}(p)$. Indeed, this inequality follows from the elementary inequality $(x+y)^\alpha\leq x^\alpha+y^\alpha$ for $x,y\geq 0$ and $\alpha\in(0,1)$.

\end{document}